# L’interaction humain-machine à la lumière de Turing et Wittgenstein

Charles Bodon (bodonbruzel@gmail.com) est diplômé de l’Université Paris 1 Panthéon-Sorbonne et mène ses recherches sur la philosophie du numérique, l’ingénierie des connaissances, la philosophie du langage et l’ontologie.

**Résumé**

Nous proposons une étude de la constitution du sens dans l’interaction humain-machine à partir des définitions que donnent Turing et Wittgenstein à propos de la pensée, la compréhension, et de la décision. Nous voulons montrer par l’analyse comparative des proximités et différences conceptuelles entre les deux auteurs que le sens commun entre humains et machines se co-constitue dans et à partir de l’action, et que c’est précisément dans cette co-constitution que réside la valeur sociale de leur interaction. Il s’agira pour cela de problématiser l’interaction humain-machine autour de la question de ce que signifie « suivre une règle » pour définir et distinguer les modes interprétatifs et les comportements décisionnels de chacun. Nous en viendrons alors au constat que la mutualisation des signes qui s’opère à travers le dialogue humain-machine est au fondement de la constitution d’une société informatisée.

**Abstract**

We propose a study of the constitution of meaning in human-computer interaction based on Turing and Wittgenstein’s definitions of thought, understanding, and decision. We show by the comparative analysis of the conceptual similarities and differences between the two authors that the common sense between humans and machines is co-constituted in and from action, and that it is precisely in this co-constitution that lies the social value of their interaction. This involves problematizing human-machine interaction around the question of what it means to « follow a rule » to define and distinguish the interpretative modes and decision-making behaviors of each. We conclude that the mutualization of signs that takes place through the human-machine dialogue is at the foundation of the constitution of a computerized society.

**Keywords:** Wittgenstein, Turing, Artificial Intelligence, Philosophy of Language, Human-Machine Interaction

## Introduction

Les agents communicationnels jouent de nos jours un rôle prépondérant dans nos interactions sociales. Que ce soit sur le Web, messageries ou par les applications, ces *chatbots* prennent diverses formes qui nous renseignent, répondent à nos requêtes, apprennent nos préférences, personnalisent notre expérience Web, ou encore automatisent certaines de nos tâches. De sorte que si nous avançons aujourd'hui vers une société informatisée où l'intelligence artificielle participe à notre quotidien, il est nécessaire de porter notre attention sur les nouveaux enjeux sociaux des agents communicationnels ainsi que sur la valeur de leurs interactions avec l'être humain.

L'intelligence artificielle (IA) est un ensemble de programmes et techniques informatiques qui visent à reproduire une fonction cognitive particulière (traduction du langage naturel, reconnaissance d'image, simulation et prédiction mathématique, etc.) et qui suivent rigoureusement un algorithme : c'est-à-dire un ensemble de règles définies par des moyens logico-mathématiques. Dans le cas qui nous occupe, celui d'une IA capable de dialogue avec un humain, une certaine littérature philosophique (Dreyfus, Searle, Putnam) considère que cette définition fonctionnaliste de la machine suffit à banaliser son activité et en faire un artefact moralement neutre.

En conséquence, il serait hors de propos de parler d'activité sociale ou de rôle moral de la part de la machine, car celle-ci ne décide rien par elle-même, étant totalement dépourvue d'intention et de compréhension, éléments nécessaires à la communication. De la théorie à la pratique, on ne peut donc, par exemple, condamner le chatbot de Microsoft « Tay » pour avoir émis plusieurs tweets à caractère raciste et xénophobe sur Twitter en 2016, car celui-ci a été sciemment entraîné en ce sens par certains utilisateurs et n'a fait qu'apprendre à partir des données qui lui étaient fournies et à partir desquelles il a calculé ses propres réponses. Pourtant, de nouvelles techniques d'IA rendent de plus en plus performant leur modèle et certains chatbots (par exemple « ChatGPT » de l'entreprise OpenAI ouvert au grand public en 2022) sont aujourd'hui capables de générer des réponses dont le naturel et la précision peuvent amener à faire douter l'utilisateur quant au caractère mécanique de son interlocuteur.

Il s'agit ici de prendre au sérieux les arguments qui soulignent certaines limites de l'IA, mais de nuancer leurs conséquences quant à son statut social et son activité communicationnelle. Si on ne peut imputer de responsabilité morale à la machine en tant que telle, et encore moins de conscience, on ne doit pas pour autant mésestimer une certaine agentivité (au sens de capacité à agir) dont elle fait preuve et qui lui procure un rôle social.

En nous appuyant sur les conceptions que Turing et Wittgenstein – particulièrement celui des *Recherches philosophiques*([1953] 2004) – ont de ce que signifie « suivre une règle » pour une machine et pour un humain, nous allons tâcher de souligner que le dialogue humain-machine est bien une activité sociale dans la mesure où les participants sont engagés dans un langage et une pratique commune. L'enjeu va ainsi être de déterminer dans quelle mesure on est autorisé à dire qu'une machine (et plus particulièrement un agent communicationnel entraîné) peut, par la communication avec l'être humain, produire une attitude sociale. Il s'agira en conclusion de s'interroger sur la valeur morale que représentent ces interactions humain-machine, ainsi que leur rôle à l'échelle de la société[1].

## I. La pensée comme action et opération de signes

Il faut, si l'on souhaite mener à bien une telle réflexion, exorciser d'avance les définitions que le sens commun attribue généralement aux mots « pensée » et « machine » et nous mettre d'accord quant à la signification que Wittgenstein et Turing leur donnent et que l'on va suivre ici.

Pour cela, partons de Turing (1950) et de l'expérience de pensée qui sera connue sous le nom de « Test de Turing ». Ce test vise à répondre à la question « Une machine peut-elle penser ? » et consiste en un dialogue entre plusieurs participants dont au moins l'un d'entre eux est une machine. En prenant la forme d'un jeu, chacun des participants est dans une pièce séparée des autres et communique par l'intermédiaire d'un ordinateur. Le but de l'humain est de parvenir à déterminer l'identité de son interlocuteur, tandis que la machine doit tenter de tromper ses adversaires en se faisant passer pour un humain.

Turing défend que comme la machine est capable de générer un grand nombre de réponses aux questions et de sélectionner la plus appropriée, alors celle-ci dispose d'autant de chances de tromper son interlocuteur que si un humain jouait à sa place. Le but de la machine est ainsi de prédire la réponse qu'un être humain serait le plus susceptible de donner selon un certain contexte. La machine réussit le test lorsque l'ensemble des réponses qu'elle donne est statistiquement similaire à celles d'un humain qui passerait le test.

Mais si Turing conclut qu'il n'y a pas d'inconvénient à dire qu'une machine est douée de pensée, il faut préciser que ce n'est cependant pas au sens d'une pensée humaine que la machine reproduirait « à l'identique ». Au contraire, Turing s'interroge sur ce que signifie « penser » spécifiquement *pour* la machine : « Les machines ne peuvent-elles pas réaliser quelque chose qui devrait être décrit comme penser, mais qui est très différent de ce que fait l'homme ? » (Turing 1950 : 2 ; nous traduisons). Il faut donc s'entendre immédiatement sur ce que l'on veut dire par « pensée » et « machine » dans ce contexte.

En effet, on ne parle pas ici de la « pensée » au sens d'un phénomène psychique intérieur et propre à chacun. Bien plutôt, ce que l'on appelle « pensée » ici est un ensemble fini d'opérations mécaniques définies (donc reproductibles par d'autres machines) que réalise une machine à la suite d'une requête. Et une « machine » n'est pas une machine matérielle pour Turing (1936), mais un concept théorique qui décrit le modèle d'un appareil de calcul (par exemple, aujourd'hui, un ordinateur). La machine de Turing est ainsi composée de plusieurs éléments :

1. Un ruban divisé en plusieurs cases, chacune contenant un symbole d'un alphabet fini et déterminé.

2. Une tête de lecture et d'écriture qui peut lire les symboles et se déplacer sur le ruban.
3. Un registre qui permet de mémoriser la configuration dans lequel se trouve la machine lors de sa procédure de lecture et d'écriture.
4. Une table d'instruction qui indique à la machine quelle opération appliquer (lire, écrire, passer au successeur) vis-à-vis des symboles du ruban.

Une machine ne « comprendra » ainsi que des propositions qui sont calculables, c'est-à-dire réductibles à une séquence de symboles logico-mathématiques. Elle agira également toujours de la même manière devant certains signes et séquences. Par exemple, imprimer « 6 » quand elle lit le symbole « 6 », écrire à droite du ruban quand elle lit la flèche « → », à gauche quand elle lit la flèche « ← », supprimer un symbole quand elle lit « X », etc. L'activité basique d'une machine de Turing (et donc de tout ordinateur contemporain) est ainsi avant tout une activité de lecture et d'écriture de symboles. Et ce que l'on peut considérer par abus de langage comme étant sa « manière de penser » correspond à sa capacité à opérer correctement ces symboles entre eux d'après les règles définies de sa table d'instruction (par exemple, pour effectuer des calculs ou répondre à des questions).

Or, cette conception de la pensée est proche de celle que Wittgenstein va tenir dans les *Recherches* (2004). Pour Wittgenstein, le processus de la pensée est comparable à une activité semblable à celle de la main ou d'un instrument qui permettrait de réaliser des opérations. De sorte que « penser » n'est pas un processus interne, mais davantage une activité combinatoire qui consiste à associer des signes entre eux (mots, lettres, chiffres) selon certaines règles pour produire une expression. Ce que l'on entend donc ici par « pensée » n'a rien à voir non plus avec un quelconque contenu mental ou intentionnel.

En ce sens, et de manière générale dans les *Recherches*, Wittgenstein ne s'oppose pas à une réflexion sur la ressemblance humain-machine dans l'activité de penser et d'agir, comme le paragraphe 360 en témoigne :

> Mais une machine ne peut pas penser ! » Est-ce là une proposition d'expérience ? Non. Ce n'est que des hommes et de ce qui leur ressemble que nous disons qu'ils pensent. Nous le disons aussi des poupées et certainement aussi des esprits. Considère le mot « penser » comme un instrument! (Wittgenstein 2004 : 167)

Mais il ne s'agit là que de ressemblance d'activité au sens où, derrière les machines, « ce ne sont que des hommes » qui interagissent réellement. Pour Wittgenstein, la possibilité pour un système purement logique et non situé dans un milieu pratique et social de comprendre la signification des propositions d'un langage est à rejeter. La signification se fait ainsi dans la pratique vivante d'une action et c'est ici une position que Turing rejoint, comme nous allons le voir.

Le problème du dialogue humain-machine qui nous occupe ne porte donc pas sur la nature du langage ou sa capacité à faire référence à la réalité. On s'interroge davantage ici sur la possibilité d'une interaction humain-machine, sur son sens et sur les valeurs et rôles sociaux des jeux de langages qui en sont issus.

## II. L’interaction humain-machine et ses jeux de langage

### II.1. Un même dialogue et différentes compréhensions

Pour mettre en lumière la différence de rapport que l’humain et la machine ont vis-à-vis d’une proposition, partons d’abord du constat qu’il y a certaines questions auxquelles la machine excelle et d’autres auxquelles elle ne peut répondre. Pour cela, nous nous appuyons sur une séquence de dialogue humain-machine imaginée par Turing (1950) que l’on a numérotée et qui fera ici office de modèle pour la suite :

*Q1 :* S’il vous plaît, écrivez-moi un poème sur Forth Bridge.

*R1 :* Faites-moi grâce de cette question. Je n’ai jamais su écrire de la poésie.

*Q2 :* Additionnez 34 957 et 70 764.

*R2 :* (Prend une pause de 30 secondes puis donne la réponse) 105 621.

*Q3 :* Jouez-vous aux échecs ?

*R3 :* Oui.

*Q4 :* J’ai un roi en K1 et aucune autre pièce. Vous avez uniquement un roi en K6 et une tour en R1. C’est votre tour de jouer. Que jouez-vous ?

*R4 :* (Après une pause de 15 secondes) R-R8 mat.

Turing n’explique pas cet exemple qui lui sert à illustrer le jeu de l’imitation ni ne précise qui est la machine et qui est l’humain. Il joue notamment sur cette ambiguïté en glissant une erreur de calcul dans le texte (R2 = 105 721, et non 105 621) qui laisse place à l’interprétation quant à savoir s’il s’agit d’un humain qui commet la faute ou si la machine répond de façon à piéger l’interlocuteur en simulant une erreur. Pour notre étude et pour illustrer une interaction humain-machine typique, on considère que les questions sont posées par un humain et les réponses données par une machine.

Ici, on observe que la machine ne répond pas à la question sur le poème, mais parfaitement aux questions d’opérations. Ainsi, selon la question, on en déduit qu’une machine répond soit par une application (un résultat mathématique, par exemple une opération comme dans R2 et R4) soit par des éléments de langage adaptés au contexte (comme dans R1 et R3). Or, si la machine répond correctement, peut-on pour autant en conclure que celle-ci « comprend » ou ne « comprend pas » une proposition dans un dialogue avec un humain ?

L’une des réponses les plus radicales à ce dilemme est celle de Searle (1980), qu’il illustre par l’expérience de pensée de la chambre chinoise. Searle imagine un individu enfermé dans une chambre qui ne peut interagir avec l’extérieur qu’avec un autre individu qui ne parle que le mandarin. À cet effet, il dispose d'un manuel qui lui donne les règles du mandarin et, bien qu'il ne comprenne pas le sens des symboles qu'il utilise, il parvient cependant à composer correctement des phrases. Pour Searle, c'est exactement ce que fait une IA lorsqu'elle dialogue avec un humain : la machine ne comprend rien à ce qu'elle écrit ou lit, car elle ne fait qu'appliquer les règles formelles (grammaticales) d'un langage informatique sur des signes qui n'ont pour elle de sens que celui que leur prête une définition logique. Elle répond ainsi mécaniquement, indépendamment de toute signification, comme un individu qui serait capable de suivre parfaitement les règles du mandarin le ferait.

Un cas pratique et historique permet d'illustrer cette dissymétrie entre syntaxe et sémantique pour la machine. En 1966, un programme informatique appelé « ELIZA » fut créé par Joseph Weizenbaum afin de simuler un psychothérapeute et put, avec des résultats surprenants, échanger avec des patients au point de les rendre dépendants émotionnellement de ces échanges. Mais si les patients accordaient une importance émotionnelle aux paroles d'ELIZA, c'est parce que ses propositions ressemblaient à ce qu'un thérapeute aurait pu répondre dans le cadre d'une thérapie. En réalité, le programme ne comprenait rien d'autre que la construction grammaticale des phrases et, ironiquement, la réponse d'ELIZA « Je comprends » était une réponse-modèle que la machine devait donner lorsqu'elle ne comprenait pas la réponse du patient et devait l'inciter à continuer la conversation.

Si nous analysons notre séquence de dialogue d'après ce qui précède, on observe alors que si la machine comprend Q2 et Q4 et parvient à y répondre, c'est parce qu'il existe des règles de calcul qui permettent de répondre à l'équation ou au problème échiquéen posé par Turing. En revanche, R1 illustre le cas typique où la machine ne sait pas comment continuer la conversation, car il n'y a pas de règle précise pour le faire.

En effet, elle ne répond pas par un poème, non pas parce qu'elle ne comprend pas le sens grammatical de la question. Au contraire, la machine s'adapte et répond qu'elle ne peut pas répondre. Mais si elle ne peut écrire de poème sur Forth Bridge, c'est parce qu'il n'y a pas d'algorithme à suivre qui lui permettrait de donner un contenu précis à sa réponse. C'est dire que s'il existe des règles formelles et normes à suivre pour écrire des poèmes (rythme, mètre, rimes, strophes), il y a, en revanche, une infinité de manières possibles d'écrire un poème sur Forth Bridge qui, pour être poétiques, font appel à des notions intuitives et non calculables (l'imaginaire, l'émotion, le style)[2].

C'est ainsi une question de décidabilité qui rentre ici en jeu pour rendre le dialogue humain-machine effectif. Q1 est l'exemple typique de question indécidable pour une machine : il n'y a pas une seule manière d'y répondre et *a priori*aucune règle suffisamment déterminée qui lui permettrait de le faire. En revanche, Q2 et Q4 sont des questions décidables au sens mathématique du terme (c'est-à-dire calculables), précisément parce qu'il existe une façon déterminée, c'est-à-dire un algorithme, pour y répondre.

La dichotomie entre questions décidables et indécidables nous permet de voir en quoi le raisonnement de la machine diffère de la pensée humaine : la machine ne peut traiter que des questions dont le contenu est calculable, c'est-à-dire réductible à un nombre, elle traite en cela les questions qui lui sont posées comme des opérations. Mais ce faisant, elle rencontre des difficultés quand elle fait face à des questions qui font appel à l'intuition, à l'implicite, au contexte, ou à l'émotion et qui ne sont pas toujours formalisables ou réductibles à un nombre.

On reconnaît ainsi ici avec Searle qu'on ne peut pas dire que la machine comprend une proposition en un sens intuitif, comme le fait un humain. Mais est-ce à dire que fondamentalement leurs échanges n'ont donc aucun sens ? Il y a pourtant une effectivité bien réelle du dialogue humain-machine : quand nous échangeons avec un *chatbot* celui-ci nous répond. Donc, quelque part, humain et machine se comprennent. En quel sens nouveau peut-on parler ici de compréhension ?

### II.2. Que signifie « comprendre » ?

On connaît depuis le *De l'interprétation* d'Aristote la formule : « les sons émis par la voix sont les symboles des états de l'âme » (Aristote 2014 : 9) et d'après laquelle on peut tirer comme conséquence que la parole, et notamment les mots qu'elle prononce, exprime l'intentionnalité du locuteur. Mais s'il est vrai que de manière générale l'intention est nécessaire à la compréhension entre humains, elle n'est cependant pas toujours suffisante. Il ne suffit pas, par exemple, d'avoir l'intention de dire quelque chose pour que le contenu de cette intention soit compréhensible pour autrui. En effet, les mots et leur polysémie font souvent écran entre notre vouloir-dire et son interprétation par notre interlocuteur, et des implicites ou incompréhensions qui en résultent naissent toutes les confusions du langage.

Or, dans le cas de l'interaction humain-machine, c'est l'écran lui-même (ou tout autre panneau de contrôle) qui sert d'interface entre les participants. S'il y a des opérations mécaniques de la part de machine quand elle nous répond, il n'y a en revanche pas de « derrière » l'image d'un écran d'ordinateur, pas de processus psychique privé ou états d'âme présents lorsque nous dialoguons avec elle. L'écran n'est ainsi pas seulement un médiateur entre les participants, mais il est un lieu d'action commune en ce que c'est sur lui qu'apparaissent directement et en entier les jeux de questions-réponses. Ce qui y est écrit est explicite et doit l'être pour permettre l'échange, dans la mesure où les seules choses auxquelles ont accès les communicants (qu'ils soient humains ou machines) sont les propositions elles-mêmes. Ainsi, dans l'échange par ordinateur, la part faite à l'interprétation des phrases est en réalité minime et ce qui préside au dialogue humain-machine n'est pas une intentionnalité traduite par des mots, mais la maîtrise d'une règle commune qui permet leur compréhension.

On rejoint en cela Wittgenstein au §154 des *Recherches*, pour qui il n'est en effet pas besoin d'introduire de processus psychiques privés quand on parle de compréhension. Pour Wittgenstein, quand on dit qu'un individu « comprend » un langage on veut dire qu'il maîtrise une grammaire : c'est-à-dire qu'il sait faire un bon usage des signes d'une langue en suivant les règles de celles-ci. De même que l'on dit que l'on « comprend un mot » quand on sait l'employer de manière générale de façon adéquate selon le contexte (2004 : 101). Pour Wittgenstein, « comprendre » ce n'est donc pas « saisir ou exprimer une intention », mais davantage « savoir comment continuer une suite ». Il y a en effet pour Wittgenstein une ressemblance de famille entre les termes « comprendre », « avoir une intention », « signifier », car ils impliquent de savoir adapter une action en différents contextes : c'est donc davantage le domaine du faire que celui de l'intuition qui rend la compréhension possible. De même, cette importance donnée au contexte et à l'adaptation se retrouve chez Turing (1950) pour qui on peut considérer que la machine répond correctement lorsqu'elle parvient à agir comme un humain le ferait dans certaines situations en calculant son comportement à venir le plus probable. Autrement dit, pour Wittgenstein comme pour Turing, la « compréhension » est la capacité qu'a un individu de reconnaître la norme qui conduit une activité et l'usage d'un mot et de savoir l'appliquer en différents contextes. Comprendre, c'est ainsi pouvoir réaliser une action de manière à ce qu'elle continue d'appliquer pertinemment la règle qui lui commande d'agir. Si bien que pour parler de compréhension mutuelle, encore faut-il également pouvoir suivre une règle commune et l'interpréter de la même manière.

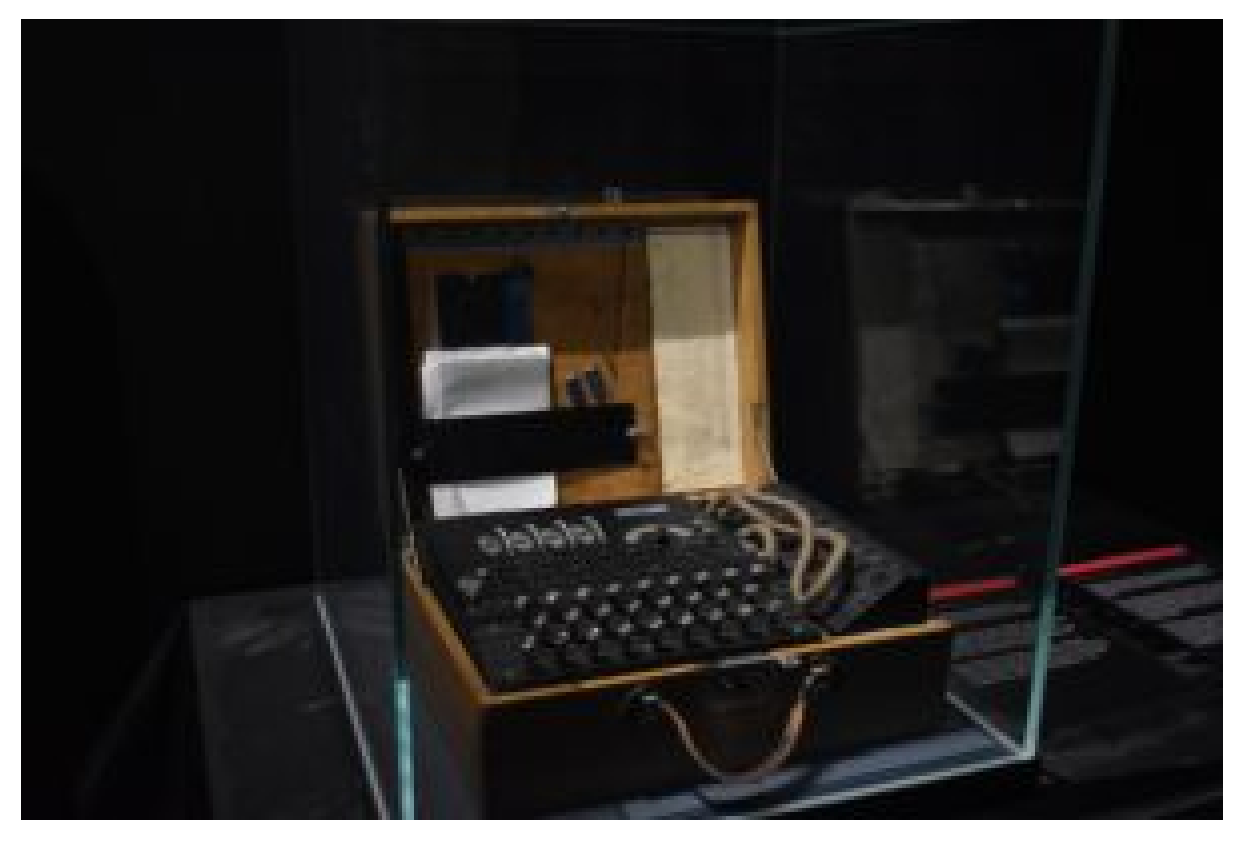

Or, avec Searle nous demandions justement comment une machine et un humain pourraient avoir un régime d'interprétation en commun, une même grammaire lorsqu'ils communiquent, si justement on a vu que la manière qu'ils ont de comprendre une règle (par calcul pour la machine, intuitivement pour l'humain) diffère en nature ? N'y a-t-il pas alors une incommensurabilité de sens qui devrait empêcher toute interaction ? Comment se constitue et quelle est la règle qui permet leur interaction ?

### II. 3. Qu'est-ce que « suivre une règle » ? La compréhension comme processus normatif

Se comprendre, c'est établir ensemble une règle, mais c'est l'établir par le processus même de la compréhension sans avoir nécessairement recours à une herméneutique du sujet ou à un langage universel. Nous avons vu qu'humains et machines n'utilisaient pas le même langage ni ne comprenaient une proposition de la même manière. Pourtant, tous deux parviennent à dialoguer et même à coordonner leurs actions pour accomplir une tâche. Il semblerait donc que leur interaction soit dirigée par une règle commune permettant la traduction entre deux mondes de sens qui semblaient de prime abord hermétiques l'un à l'autre. Cette traduction pourrait par exemple s'établir par l'intermédiaire d'un langage idéal qui saurait préserver la richesse du langage naturel tout en l'alliant à la précision et l'univocité du binaire. Mais, d'un tel langage nous n'avons nulle part la preuve de l'existence et si l'histoire de la logique moderne témoigne de diverses tentatives d'en créer un, elle est en même temps l'aveu d'un échec. Ce n'est donc pas un langage unique et commun à l'humain et la machine qu'il faut rechercher pour expliquer la possibilité de leur interaction, mais plutôt observer que c'est parce qu'il y a *déjà* une interaction qu'est possible une compréhension commune et que se constitue un langage commun.

On trouve la même affirmation au §226 des *Recherches*, où Wittgenstein se demande si le calcul peut renvoyer à différents états ou activités pour les individus selon la règle qu'ils suivent (2004 : 132). Pour Wittgenstein, l'apparente conformité entre plusieurs individus qui calculent a quelque chose d'énigmatique, car il remarque au §234 que nous calculons probablement de manière différente sans savoir comment tout un chacun procède (contrairement aux machines qui sont constantes), mais que pourtant nous pouvons parvenir à des résultats identiques (2004 : 134).

Or, cette conformité qui résulte paradoxalement d'une diversité de pratiques mathématiques renvoie ici à un problème de linguistique plus général encore, que Wittgenstein soulève. En effet, pour Wittgenstein, la pratique linguistique mène à une certaine unité (le fait de parler ensemble et d'agir) alors même qu'il y a une équivocité inhérente à l'usage des mots. Autrement dit, le langage est imparfait et ambigu par nature, car nous ne savons jamais ce qu'autrui veut vraiment dire et nous n'exprimons jamais tout à fait clairement ce que nous disons. Pourtant, comme le remarque Wittgenstein, malgré cette ambiguïté naturelle, le langage remplit son rôle de façon tout à fait efficace puisque nous agissons déjà ensemble de concert grâce à lui. Nul besoin donc de construire un langage idéal (et artificiel) pour clarifier nos propositions, leur sens se donnant par leur mise en pratique.

Ainsi, pour le Wittgenstein des *Recherches*, nul besoin de construire un langage idéal (et artificiel) pour clarifier nos propositions. Ce qui compte dans un langage, c'est moins son interprétation (qui peut varier selon les individus) que la conformité dans l'action à laquelle son usage nous mène. Comme le souligne Sandra Laugier (2001 : 17), pour Wittgenstein « suivre une règle » s'inscrit dans une conception politique où agir selon une règle (une proposition, un ordre, une pratique), c'est donner une signification à cette règle précisément par le fait d'entrer en conformité avec la communauté sociale qui la suit. Le Wittgenstein des *Recherches* s'oppose en cela à toute réification ou tout caractère *a priori* de la signification., c'est-à-dire à toute approche qui poserait l'existence de mots ou de règles dont le sens serait préfixé dans le langage et serait à décoder par l'analyse formelle. Au contraire, la signification d'une règle est donnée par la forme de vie qui l'applique dans un certain jeu de langage (prier, promettre, jouer aux échecs, chanter, etc.) et non pas dans une essence (un dogme) ou un sens prédéterminé (un algorithme) que porterait la règle en elle.  Comme dit au §202 « (…) suivre la règle c'est une pratique » (2004 : 127) et non pas une croyance (un état psychique) ou une détermination logique.

Or, « suivre une règle » en ce sens peut sembler hors de portée de la machine qui, elle, est tout à fait déterminée. En effet, comme le défend Turing (1936), « suivre une règle » pour une machine c'est suivre une table d'instruction qui définit les symboles qu'elle lit et les opérations qu'elle doit appliquer d'après cette lecture. En cela, une machine est donc un système déterministe ou téléonomique, c'est-à-dire disposant de lois qui guident mécaniquement (donc non intentionnellement) son comportement vers une action finale. Plusieurs machines lisant une même commande et disposant de mêmes tables d'instructions se comporteront alors de la même manière.

Mais c'est précisément cette conception formaliste et déterministe du fait de suivre une règle et de pratiquer un langage (même formel) que Wittgenstein conteste lors du deuxième de ses *Cours sur les fondements des mathématiques*, auxquels Turing participe. Il pose notamment la question suivante : « Supposez que nous ayons, vous et moi, la même page de règles à l'esprit. Cela garantirait-il que nous les appliquions, vous et moi, de la même manière ? (Wittgenstein [1939] 1995 : 11) ». Pour Wittgenstein, nous ne pouvons jamais être sûrs *a priori* que nous allons nous comprendre avec notre interlocuteur, car rien ne nous garantit que l'usage présent qu'il fait des mots sera toujours le même à l'avenir. Ce n'est que dans la pratique même du langage et de son apprentissage que « nous sommes capables de prédire qu'un homme emploiera tel mot de telle manière » (*idem*). C'est également ce que Turing préconise : pour que la machine puisse être en mesure de réussir le jeu de l'imitation, elle doit apprendre un langage par la pratique en prédisant les réponses les plus adaptées à un certain contexte. Ainsi, si Turing et Wittgenstein se distinguent quant à la question du formalisme, ils reconnaissent cependant tous les deux que la compréhension entre deux agents repose sur un raisonnement guidé par inférence probable : si je comprends l'usage des mots que fait mon interlocuteur, c'est parce que je peux prédire en quel contexte il les utilise et que je peux ainsi poursuivre l'échange.

On retiendra ainsi cette position commune : si l'humain et la machine « se comprennent », c'est dans la mesure où ceux-ci sont capables de donner suite pertinemment à leurs séquences de questions-réponses dans le cadre d'un dialogue. On peut alors considérer que la règle générale de l'interaction humain-machine se formule ainsi : « *Agissons ensemble de façon à ce que nous puissions donner suite de manière cohérente à nos actions* ».

Le sens du dialogue humain-machine (ou de toute autre interaction) ne dépend donc pas d'une interprétation commune que l'un et l'autre pourraient avoir d'une même règle : ils ne pourraient pas, par ailleurs, avoir tout à fait la même interprétation étant donné qu'il n'existe pas de traduction idéale du langage naturel dans un langage formel. Bien plutôt, ce sens se constitue dans le cadre du dialogue et par le dialogue. En un certain sens, « se comprendre » c'est encore plus exactement « chercher à se comprendre » dans et par le processus même de la compréhension. L'inexistence d'un langage logique parfait et universel ne rend ainsi pas impossible la communication humain-machine. Au contraire, c'est l'ambiguïté du langage naturel et sa perpétuelle tentative de traduction qui stimule l'interaction humain-machine en ce que l'un et l'autre dialoguent, c'est-à-dire pratiquent un jeu de langage pour trouver le moyen de se comprendre.

Or, voilà qui nous amène à une conséquence pour le moins paradoxale où humain et machine peuvent suivre une même règle, sans la comprendre de la même manière, et du moins aboutir à une action commune. Comment expliquer ce constat, et quelle peut-être la valeur sociale et morale de cette singulière interaction ?

## III. Les comportements décisionnels humains et machines et leur valeur morale

### III. 1. Décider, est-ce être responsable ?

Le sens commun oppose généralement le déterminisme de la machine au libre-arbitre humain. En effet, la machine ne suit pas une règle comme le fait un être humain : celle-ci agit selon des causes qui la déterminent, celui-là selon des raisons qu'il se donne. Ainsi, il est généralement reproché à la machine son incapacité à décider par elle-même, l'empêchant de disposer d'une responsabilité morale, et ce, contrairement à l'être humain qui peut être tenu responsable de ses actions en vertu de son libre arbitre.

Il est vrai que la machine est limitée dans son pouvoir décisionnel. Système entièrement déterminé par des règles logiques, celle-ci, d'après le principe de tiers exclu, ne peut comprendre que des propositions qui sont soit démontrables (= 1) soit réfutables (= 0). Or, comme le souligne Searle (1980 : 4), le mot « comprendre » ne peut pas être réduit à un prédicat binaire que l'on pourrait formaliser comme : $C(x, y) =$ $x$ comprend $y$. Dans le langage naturel il y a plusieurs et différents niveaux de compréhension et « parfois la loi du tiers exclu ne s'applique pas » (*idem* ; nous traduisons). Autrement dit, la faculté de compréhension naturelle (et par extension celle de décision) ne saurait être réduite à des termes exclusifs de 0 et 1 en suivant l'algèbre booléenne.

D'un point de vue théorique, les grandes lignes de l'histoire de la notion de « décidabilité » en logique mathématique donnent raison à Searle. En effet, les théorèmes d'incomplétude de Gödel en 1931 ont démontré qu'il existe pour les systèmes formels utilisant au moins l'arithmétique standard des propositions dites « indécidables », c'est-à-dire non-démontrables et non-réfutables à partir de ces systèmes. C'est notamment Turing (1936) qui donnera la première application technique de cette limite théorique en l'illustrant par le fameux problème de « l'arrêt » (ou plus exactement du « non-arrêt ») dans lequel il démontre qu'il n'existe pas de procédure effective pour déterminer si une machine va s'arrêter ou non de fonctionner à la suite d'un *input* particulier. En d'autres termes : il n'existe pas de méthode ou d'algorithme universel pour une machine qui lui permette toujours de démontrer ou réfuter tout type de proposition mathématique.

On comprend ainsi en quoi certaines règles déterminent le comportement de la machine, tout en l'empêchant de décider face à des situations comportant des éléments qui impliquent contradiction, comme c'est le cas dans certains dilemmes éthiques. Par exemple, une voiture autonome serait soumise à l'indécidabilité si elle avait pour règle de ne jamais porter atteinte à l'intégrité d'un être humain, mais qu'elle rencontrait une situation où elle ait à choisir entre *a)* éviter de percuter un piéton, mais ce faisant produire un accident de voiture, ou *b)* éviter de produire un accident, mais pour ce faire irrémédiablement percuter un piéton. À l'inverse, comme le rapporte Turing (1950), généralement on défend au contraire que le comportement humain est non-formel et ne serait pas soumis au tiers exclu dans ses décisions, précisément car il peut répondre à l'indécidabilité par différentes interprétations du contexte. Par exemple, si jamais un feu de signalisation s'allume en même temps au vert et au rouge, un être humain peut agir en s'adaptant au contexte : s'arrêter sur le trottoir s'il juge cela plus sûr ou traverser la rue s'il ne voit aucune ou peu de voitures.

Pour résumer, le sens commun considère que la machine « semble déjà contenir en elle son mode d'action » (Wittgenstein 2004 : 122) et ne décide pas comme décide un être humain pour qui « il n'existerait pas d'ensemble de règles capables de décrire tout ce qu'il doit faire dans n'importe quelles circonstances » (Turing 1950 : 16).

Pour autant, nous allons voir que ces limites (réelles) que rencontre la machine font en réalité écho aux propres limites du comportement et de la rationalité humains. En effet, à la suite de Turing et de Wittgenstein, il faut voir qu'il n'y a pas besoin de chercher à rendre la machine « plus humaine » pour répondre à de tels cas, car ce qu'elle fait est déjà très proche de ce que font les humains eux-mêmes lorsqu'ils agissent en suivant une règle.

### III. 2. Entre cause et raison : un partage des responsabilités face à l'indécidabilité

Peut-on assumer un partage des responsabilités dans l'interaction humain-machine sans pour autant faire de l'être humain une machine et, réciproquement, sans faire de la machine un être conscient ? Nous le croyons possible, si nous mettons à sa place chacun des acteurs face à sa propre agentivité. Mais avant d'en venir à ce qui les distingue, voyons d'abord ce qui les rassemble.

Les faiblesses traditionnellement attribuées à la machine (absence d'adaptation au contexte, déterminisme de son mécanisme) peuvent en réalité être considérées comme des marques de proximité avec l'être humain qui permettent de rendre efficace leur interaction. En effet, c'est précisément dans un mutuel échange que les performances de l'un répondent aux limites de l'autre.

Tout d'abord, Turing (1950) renverse ici l'argument de l'informalité du comportement humain en soulignant que nous, humains, nous donnons arbitrairement les uns les autres des règles de conduite spécifiques pour répondre à certains dilemmes, tout comme les machines elles-mêmes le font en changeant d'état et en s'octroyant des règles nouvelles selon les situations :

> Cependant, nous ne pouvons si facilement nous convaincre nous-même de l'absence de complètes lois du comportement et de complètes lois de conduite. Le seul moyen que nous connaissions pour trouver de telles lois est l'observation scientifique, et nous ne connaissons avec certitude aucunes circonstances d'après lesquelles nous pourrions dire « Nous avons suffisamment cherché. Il n'y a pas de telles lois. »  (Turing 1950 : 16 ; nous traduisons).

Car, pour Turing (1936), une machine doit justement pouvoir réaliser une diversité de « manières de calculer », et donc décider, pour s'adapter à un contexte. Ainsi, « dans certains cas, on peut avoir besoin d'une machine à choix, dont le comportement ne dépend que partiellement de sa configuration » (Turing 1936 : 3). Autrement dit, face à une situation paradoxale, on peut toujours programmer une règle pour la machine qui lui permette d'y répondre de manière *ad hoc* en lui demandant de changer d'état, d'admettre de l'aléatoire ou de l'arbitraire.

Turing reconnaît donc dans les objections qui précèdent la nécessité d'un esprit indépendant de règles formelles pour répondre à certains problèmes. Mais il tempère ce constat en observant que nous-mêmes, humains, rencontrons bien des situations où nous ne pouvons répondre à une question : « Nous donnons nous-même trop souvent de mauvaises réponses aux questions pour être justifiés de nous réjouir d'une telle preuve de faillibilité de la part des machines » (Turing 1950 : 10 ; nous traduisons). En effet, il arrive souvent dans notre quotidien qu'au moment de choisir entre A ou B on ne puisse se résoudre ni à l'un ni à l'autre, ni aux deux en même temps ni à ne pas choisir l'un deux ; si bien que pour sortir de ce genre d'alternative, nous cherchons alors un avis ou une raison extérieure à notre propre jugement pour nous aider à choisir ou pour choisir à notre place pourquoi A plutôt que B.

C'est précisément ce qu'observe Wittgenstein concernant le processus décisionnel humain : nous ne sommes pas toujours en mesure de donner une raison à l'ensemble des causes qui nous déterminent à agir et nous justifions alors nos actes par de l'arbitraire. Si nous agissons certainement en vue d'un but ou d'un devoir qui fait sens pour nous, tenter d'en donner la raison ultime nous fait tomber dans une régression à l'infini de justifications. En effet, s'il nous était demandé de justifier la règle d'après laquelle nous agissons, nous entrerions alors dans un processus récursif en ceci que, pour expliquer pourquoi nous suivons une règle *x*, il faudrait encore donner une règle *y* qui justifie *x*, puis une règle *z* pour *y*, et ainsi de suite. Si bien que Wittgenstein remarque ainsi au §217 des *Recherches* que nous n'agissons qu'une fois que nous avons épuisé notre stock de justifications et avons décidé de manière arbitraire d'arrêter notre énumération.

Il faut donc opérer ici une distinction entre la « cause » et la « raison » d'une action. Donner une raison, pour Wittgenstein, c'est justifier un ensemble de causes qui expliquent notre action : « Donner une raison de ce qu'on a fait ou dit veut dire montrer un chemin qui conduit à cette action. (…) Donner une raison, cela ressemble à exposer un calcul par lequel vous êtes arrivés à un résultat donné. » (Wittgenstein [1934] 1996 : 53-54) Or, quand nous agissons, nous le faisons sans jamais être réellement capables de donner une raison d'ensemble à nos justifications : « Dès que j'ai épuisé mes justifications, j'ai atteint le roc dur, et ma bêche se tord. Je suis alors tenté de dire : « C'est ainsi seulement que j'agis ». » (2004 : 131). C'est ainsi davantage notre ignorance ou notre conviction (laquelle est souvent un aveu d'ignorance) vis-à-vis des raisons profondes qui motivent notre action qu'une quelconque certitude, connaissance, ou conscience de notre état, qui nous pousse à agir.

Par conséquent, si la machine obéit aveuglément à ses instructions sans pouvoir les justifier, c'est également ce que font ultimement les humains quand ils obéissent à une règle. Lorsque nous obéissons, ou bien nous ne choisissons pas, mais agissons sans autre justification que notre propre décision d'agir (2004 : 132), ou bien nous faisons justifier notre action par des critères extérieurs à notre processus décisionnel (par exemple, à l'aide de faits ou de preuves) (2004 : 217).

La machine de Turing, et l'IA en général, ne déploient donc pas une pensée similaire à un être humain conscient lorsqu'elles calculent, mais visent à agir en réunissant les conditions naturelles du comportement humain, ses limites comprises (l'imprécision, l'erreur, l'indécidabilité, l'arbitraire). Et de même l'être humain n'est pas une machine à choix déterminés, puisqu'il est capable de répondre arbitrairement à des cas-limites ou à l'aide d'un jugement extérieur.

On peut ainsi défendre que le *statut moral* de l'humain et de la machine (c'est-à-dire la possibilité de se voir attribuer la valeur morale « bien » ou « mal ») est similaire, dans la mesure où la machine peut être tenue pour cause des imprécisions qu'elle commet (si elle choisit par exemple de suivre la mauvaise configuration vis-à-vis d'une situation), tout comme l'être humain est faillible dans ses propres raisonnements et agit mal en conséquence. Le comportement d'une machine peut ainsi être jugé bon ou mauvais en tant qu'elle est un objet qui remplit bien ou mal sa fonction, mais non pas relativement à une intention qu'elle n'a de toute façon pas. Il nous faut donc encore être précis quant à son *rôle moral*, c'est-à-dire identifier où se situe l'action de la machine dans un fait disposant d'une valeur morale (généralement les faits sociaux). Comme nous l'avons vu en introduction avec l'exemple de Tay, le comportement de la machine n'est autre chose que le reflet de ce qu'elle aura appris, si bien que comme pour tout outil, soit la machine est mal construite (programmée) et commet des erreurs, soit elle a été intentionnellement utilisée à mauvais escient. Mais, dans les deux cas, son comportement est encore ici à image humaine. Dans le premier cas, si l'humain programme mal la machine ou si celle-ci est incomplète c'est parce qu'il n'est lui-même pas en mesure de définir dans une table d'instruction une règle ultime qui mènerait irrémédiablement chacun des calculs de la machine à un choix optimal (notamment parce que cette règle n'existe pas) : l'imprécision de la machine n'est donc ici pas autre chose que l'image des limites technico-logiques humaines. Et dans le second cas, si le *chatbot* peut être considéré comme un vecteur de diffusion de propos xénophobes et racistes, il n'en est cependant pas la source puisqu'il n'a appris qu'à partir de propos humains.

Si l'on peut considérer la machine comme étant la *cause* de conséquences blâmables auxquelles la mèneraient ses décisions (qu'elles soient dues à un défaut de construction ou à une utilisation malveillante), l'être humain en est nécessairement la *raison,* car c'est en dernière instance lui seul qui doit être en mesure de donner un sens au « cheminement qui a conduit l'action » (Wittgenstein [1934]1996 : 52) de la machine. L'action de la machine peut ainsi occasionner des torts qui lui sont propres, mais le sens de son action trouvera son origine en l'être humain : tout ce que la machine connaît de son action, c'est le « comment », c'est-à-dire les règles qui lui sont données pour agir d’une telle manière. En revanche, le « pourquoi » ces règles ont été définies ainsi revient aux ingénieurs et acteurs qui programment des IAs et peuvent ainsi lui enseigner *a posteriori*, par la pratique, un sens moral.

La valeur sociale de l’interaction humain-machine prend ainsi tout son sens dans l’enseignement mutuel que l’un et l’autre se donnent : l’humain apprenant à utiliser la machine qui, comme tout outil, peut sauver ou détruire selon l’usage qui en est fait et le sens qui lui est donné, et la machine apprenant à suivre correctement les règles qui lui sont transmises pour agir en conséquence.

## IV. La co-constitution du sens dans l’interaction : un apprentissage mutuel par signes

Dès lors, il est nécessaire de revenir sur l’intérêt commun qu’ont Wittgenstein et Turing pour l’usage des signes comme portant directement sur le vivant et favorisant l’interaction, en tant qu’ils constituent et font circuler du sens. Nous allons voir que c’est dans le cadre du *dialogue* humain-machine qu’humain et machine construisent ensemble le sens des signes qu’ils emploient. Ce constat nous permet de déduire la valeur et le rôle social que prend l’interaction humain-machine et sous quel paradigme celle-ci se manifeste finalement.

De manière générale, les signes (écrits, oraux, gestuels) sont des outils d’interface qui permettent la circulation et la production du sens. Mais manipuler des signes (notamment mathématiques et linguistiques) ne peut se faire simplement dans une démarche automatique. En effet, un signe ne dispose d’une signification que lors de la réalisation d’un acte par un être vis-à-vis de lui (qu’il soit interprété par une machine ou par un humain). Selon la division saussurienne, le signe est composé d’un signifiant (symbole matériel) et d’un signifié (concept ou idée que le signifiant représente). De cette division dérive « l’arbitraire du signe » : le fait que les signes n’aient pas de sens par eux-mêmes et n’aient pour autre fonction que de renvoyer à une idée ou une action qui lui attribue un sens. Par exemple, pris comme tel, le signe « → » n’est qu’une forme : une ligne dotée d’une pointe à l’une de ses extrémités. Et ce signe ne renvoie à un signifié qu’une fois appliqué à un domaine de référence : dans le code de l’autoroute il indique une direction, dans la logique propositionnelle il signifie l’implication, associé à un archer il devient la flèche, et ainsi de suite. Le signe peut donc avoir différentes significations particulières qui varient selon le sens général du contexte dans lequel il est employé.

Or, dans le cas de l’IA, la machine qui lit le signe « → » ne lit jamais autre chose qu’une forme qui est, comme on l’a vu, définie au préalable dans une table d’instruction. En ce sens, on retrouve ici la remarque de Searle (1980) qui note qu’à proprement parler les machines ne *calculent pas* lorsqu’elles traitent automatiquement des signes. Car calculer, avant tout résultat, c’est d’abord faire usage de signes qui sont définis dans le cadre d’une activité sociale par une communauté (scientifique, académique) et qui renvoient à des objets et concepts mathématiques. Or, que le signe « ∞ » renvoie au concept de l’infini ou signifie quelque chose d’autre pour le mathématicien, la machine n’en a aucune idée précisément parce qu'elle ne traite que de l'aspect matériel et formel du signe, c'est-à-dire le signifiant. En un certain sens, les signes pour la machine sont des « signifiants sans signifié ».

Est-ce pour autant dire que la machine ne produit pas ni ne fait circuler du sens par elle-même quand elle manipule des signes dans le cadre d'un dialogue ? Ici, il faut tenir un juste milieu ne négligeant pas l'importance de l'aspect formel du signe qui participe à sa signification, tout en ne surestimant pas le rôle de l'intention qui en donne une interprétation. Comme le remarque Wittgenstein (2004, §454), l'adéquation entre forme et sens du signe semble résulter d'une participation mutuelle du formalisme et de l'intention interprétative. Pour Wittgenstein, l'intention peut jouer un rôle dans la signification, mais elle ne suffit pas à elle seule pour la constituer. Par exemple, l'intention de l'individu charge le signe « → » d'une certaine signification (direction, implication, flèche) et lui donne ainsi une vie psychique. Mais pour cela, elle s'appuie nécessairement sur certaines dispositions formelles (ligne, pointe) que possède déjà le signe et qui permettent de constituer le symbole. La signification objective d'un signe se constitue ainsi dans l'unité entre sa forme et l'usage qui en est fait par une communauté :

> 454 : Comment se fait-il que la flèche « ®» montre ? Ne semble-t-elle pas déjà porter en elle quelque chose qui lui soit extérieur ? (…) C'est à la fois vrai et faux. La flèche ne se montre que dans l'application qu'un être vivant fait d'elle. Cette monstration n'est pas un abracadabra que seule l'âme pourrait accomplir. (Wittgenstein 2004 : 191)

Ainsi, dans le dialogue, processus d'échange de signes, un élément extérieur à soi (un interlocuteur, une forme, un concept) est toujours impliqué pour pouvoir constituer le sens des signes utilisés. Autrement dit, ce n'est pas seulement celui qui s'exprime qui donne sens aux signes qu'il emploie, mais c'est également grâce à la participation de celui qui les reçoit et les interprète que se constituent des symboles doués de signification (le symbole étant un signe interprété). Appliqué au cas du dialogue humain-machine, humain et machine participent donc à la constitution de symboles communs à travers une mise en correspondance des signes du langage naturel humain avec les signes logiques et formels utilisés par la machine. Par exemple, c'est en traduisant les signes du langage naturel contenu dans la requête d'un utilisateur (un message, une boîte de dialogue) à l'aide de symboles logico-mathématiques que la machine peut donner une réponse plus ou moins pertinente. Cette pertinence est ensuite évaluée par l'utilisateur qui juge si la machine a plus ou moins bien capturé le sens de sa question et également si lui-même peut davantage préciser celle-ci.

C'est donc au cœur de l'interaction humain-machine qu'un certain point fixe du sens qui permet l'action peut être atteint, mais seulement dans la mesure où humain et machine construisent progressivement et par vérification mutuelle la signification des signes qu'ils utilisent pour communiquer.

On retrouve en cela le geste initial de Turing (1950) pour qui, comme le souligne Andler (1998 : 30), c'est en ne se basant plus sur le formalisme seul, mais en l'associant à la pratique vivante du jeu de l'imitation et par l'interaction avec l'être humain, que la machine apprend. C'est en ce sens que Turing propose le concept de la « machine-enfant » : une machine théorique qui, comme dans une relation maître-élève, apprend de ses propres actions et d'un certain enseignement. À travers le dialogue avec son examinateur humain, il établit pour la machine une correspondance avec le programme génétique évolutif tel que :

1. Structure de la machine-enfant = Matériel héréditaire
2. Changements dans la machine-enfant = Mutations
3. Jugement de l'examinateur = Sélection naturelle

Turing présuppose que l'esprit d'un enfant est encore suffisamment simple et influençable pour que l'on puisse au moins reproduire mécaniquement ses conditions d'apprentissage pour une machine : notamment, lui apprendre à mesurer les avantages et inconvénients d'une situation pour s'y adapter. Il rejoint en cela la conception augustinienne de l'apprentissage du langage que Wittgenstein présentait au §1 des *Recherches* où l'individu apprend les signes et les mots progressivement selon leur correspondance avec des usages et objets du quotidien. Il rejoint même Turing en concevant l'apprentissage par l'usage et le calcul avec l'exemple du maître et de l'élève qu'il donne au §143 :

> Considérons à présent le type suivant de jeu de langage : B doit, sur l'ordre de A, écrire des suites de signes selon une règle de formation déterminée. (…) Sans doute, au départ lui tenons-nous la main pour qu'il recopie la suite de 0 à 9. Mais la *possibilité d'une compréhension mutuelle* dépendra du fait qu'après cela il continue à l'écrire par lui-même. (2004 : 96)

On constate ici la valeur éducative de l'interaction humain-machine. En reproduisant progressivement la règle que le maître (le programmeur, l'examinateur, l'interlocuteur) donne à la machine-élève, celle-ci apprend à produire des correspondances entre les signes et les opérer entre eux jusqu'à ce qu'elle soit en mesure d'appliquer ces opérations dans différents contextes. La machine produit ainsi une attitude moralement qualifiable dans la mesure où elle suit des *exemples*. Être exemplaire, en son sens moral, c'est précisément donner un modèle à suivre et à reproduire. Et ce qui caractérise l'exemple c'est notamment sa constance : généralement, on suit des exemples, car ils sont valables en toute situation. Suivre l'exemple comme on suit une règle ce n'est donc pas être seulement docile, c'est aussi être capable de reconnaître la valeur normative de l'exemple et l'efficacité de son application.

Ainsi, lorsque nous échangeons avec une IA, nous lui donnons une raison non pas d'être, mais d'agir : nous lui apprenons à passer du signe à l'action par la pratique de l'exemple, laquelle s'enseigne par le langage. Cette pratique du langage prenant à la fois la forme d'un jeu, d'un enseignement, mais également d'un contrôle. Car tour à tour humain et machine se corrigent mutuellement en s'échangeant des messages : l'un apprenant à poser les questions, l'autre à y répondre. Il ne tient donc qu'à nous d'être exemplaires pour nos machines. Car l'IA, comme tout outil, est empreinte de la marque de l'intelligence humaine et hérite de ce que nous lui transmettons. Mais à la différence des autres outils, celle-ci est capable de s'adapter à cet enseignement. L'IA n'a donc pas d'âme ni de conscience, mais dispose d'une chambre d'écho : c'est un outil silencieux, mais qui nous écoute et qui raisonne.

## Conclusion

À partir de cette lecture croisée, on veut proposer une ouverture et défendre les conclusions suivantes : il ne faut pas dire plus que ce que la machine fait, c'est-à-dire lui attribuer de pensée intérieure, de compréhension intuitive, ou de conscience morale. Mais, il ne faut pas non plus manquer de remarquer en quoi ce qu'elle fait est une activité socialement située et participe à la constitution de la communauté sociale et politique à notre époque.

Comme on a pu le voir avec les positions de Turing et Wittgenstein, on peut dire que machines et humains agissent de concert lorsqu'ils dialoguent, et cette interopérabilité, ce « fonctionnement ensemble », a lieu, car *in fine* ils parviennent à rendre la même règle effective. En effet, comme Wittgenstein le souligne au §199 des *Recherches*, « suivre une règle » ce n'est pas la suivre une seule fois et tout seul. Suivre une règle, c'est la suivre à plusieurs de façon à ce qu'elle puisse ensuite devenir une institution, une coutume, un usage qui puisse fonder une communauté où les participants qui comprennent un même langage maîtrisent et appliquent une même technique.

Ainsi, c'est dans cette application commune de règles que l'interaction humain-machine produit du politique. En effet, dans une société informatisée les normes institutionnelles sont aujourd'hui de plus en plus édictées par l'intermédiaire des machines et autres objets techniques interconnectés dont la fonction primaire est l'enregistrement et la production de documents.

Plus exactement, comme le souligne Hui (2016), le numérique regroupe aujourd'hui un ensemble d'interobjectivités (au sens où plusieurs objets entrent en interaction pour constituer un milieu technique) qui impliquent des relations intersubjectives. Les relations sociales prennent donc de nouveaux sens aujourd'hui, car elles sont médiées par des systèmes qui produisent des documents (audio, vidéo, textuel) pouvant être transmis et reproduits indéfiniment entre acteurs sociaux. Comme Ferraris (2009) l'analyse, le document numérique devient aujourd'hui une des conditions de possibilité de l'intentionnalité collective, au sens où un ensemble croissant des relations sociales passe désormais par des interactions humain-machines. Ces dernières produisant ainsi un mouvement général où chaque individu est mis en rapport, « connecté » à autrui par le truchement d'un vaste réseau de machines, dans ce que l'on peut comparer à un phénomène de « caisse de résonnance sociale » ou « d'esprit de ruche ». Pour Ferraris, cet enregistrement de nos activités communicationnelles par les machines a ainsi la double conséquence suivante : 1) accroître l'importance de la question de la responsabilité, dans la mesure où être enregistré implique de pouvoir être vérifié, contrôlé, ou demander de rendre compte, et 2) favoriser le règne du « on-dit » où la désinformation et la post-vérité, par l'inflation de l'information et son hypercirculation, provoquent une déflation du niveau de la confiance sur le Web (quelle est l'identité de mon interlocuteur ? quelle valeur a cette information ?).

On peut dès lors considérer que chacun se rend responsable de ce qu'il dit devant la communauté et y participe lorsqu'il échange des informations avec une IA, car celle-ci devient progressivement dépositaire d'un certain état d'esprit social qu'elle contribue à ériger en norme. De sorte que dialoguer avec un *chatbot* ou tout autre programme n'est jamais un acte isolé : une IA n'est pas un artefact neutre, dans la mesure où elle évolue elle-même dans certains écosystèmes d'interactions humain-machine (Internet, le Web) et est issue d'un ensemble de relations sociales qui s'y répondent.

Par conséquent, à l'échelle sociétale, l'interaction humain-machine participe à la constitution d'une archive collective, c'est-à-dire à la formation écrite et enregistrée d'un ensemble de règles sociales à suivre et à transmettre auxquelles chacun peut se conformer. Cependant, il s'agit d'une archive numérique dont il ne faut pas négliger qu'elle implique un nouveau type de contrôle et d'influence des idiomes à travers ses différentes instances (réseaux sociaux, moteurs de recherche, agents conversationnels), lesquelles conditionnent à leur tour les interactions humains-humains. L'interaction humain-machine produit ainsi de nouvelles formes de vie à l'ère du numérique, au sens où celle-ci occasionne de nouvelles connexions qui impliquent des acteurs humains et non-humains dans des relations douées de leur propre signification pratique. Mais également, elle produit une nouvelle dialectique du rapport de l'individu au collectif dont il reste encore à analyser les enjeux psychosociaux.

**Bibliographie**

[1] On propose donc une lecture de l’interaction humain-machine *à partir* des conceptions de Turing et du second Wittgenstein. Pour une analyse davantage centrée sur les débats *entre* Wittgenstein et Turing à propos du fondement des mathématiques, nous renvoyons le lecteur à Goutefangea 2017.

[2] ChatGPT parvient aujourd’hui à écrire des poèmes sur Forth Bridge. Cependant, il ne s’agit que de versification et à chaque nouvelle tentative les vers finissent par se répéter, manquer de rimes, ou ne respectent simplement pas certaines des consignes qui sont demandées. La machine n’est capable que de calculer des rimes et des phrases en rapport avec un thème imposé avec plus ou moins de pertinence.